# Future of Code with Generative AI: Transparency and Safety in the Era of AI-Generated Software

By David Hanson Ph.D.

## Abstract

As artificial intelligence (AI) becomes increasingly integrated into software development processes, the prevalence and sophistication of AI-generated code continue to expand rapidly. This study addresses the critical need for transparency and safety in AI-generated code by examining the current landscape, identifying potential risks, and exploring future implications. We analyze market opportunities for detecting AI-generated code, discuss the challenges associated with managing increasing complexity, and propose solutions to enhance transparency and functionality analysis. Furthermore, this study investigates the long-term implications of AI-generated code, including its potential role in the development of artificial general intelligence and its impact on human–AI interaction. In conclusion, we emphasize the importance of proactive measures for ensuring the responsible development and deployment of AI in software engineering.

## Introduction

The integration of artificial intelligence (AI) into software development processes marks a pivotal development in the evolution of computer programming. As AI-generated code becomes increasingly prevalent and sophisticated, it introduces both significant opportunities and complex challenges for software engineering. This study investigates the multifaceted implications of AI-generated code, focusing on the critical need for transparency, safety measures, and ethical considerations in this rapidly evolving field.

The advent of AI in code generation is transforming the software development lifecycle with potential improvements in efficiency, productivity, and innovation. Nevertheless, this transformation introduces new complexities and risks that require careful consideration. As we navigate this emerging landscape, issues such as code provenance, functionality, and the possibility of concealed features have emerged as paramount concerns for developers, organizations, and society at scale.

This study aims to comprehensively analyze the current state and future trajectory of AI-generated code, with emphasis on the importance of transparency and safety measures. We investigate market opportunities for detecting AI-generated code, explore the challenges associated with managing increasing complexity in software systems, and propose potential solutions for enhancing transparency and functionality analysis. Furthermore, we explore the long-term implications of AI-generated code, including its potential role in the development of artificial general intelligence (AGI) and its impact on human–AI interaction.

By addressing these critical issues, this study contributes to the ongoing dialogue on the responsible development and deployment of AI in software engineering. Moreover, it highlights the importance of proactive measures for ensuring safety, reliability, and ethical use of AI-generated code in the future of computing.

**Current Landscape of AI-Generated Code**

The integration of AI into software development processes has increased significantly in recent years. A GitHub survey (2023) revealed that 92% of developers utilize AI-powered coding tools in their workflow, with 70% employing these tools daily. This widespread adoption is primarily driven by the potential for enhanced productivity and efficiency in code generation.

However, the deployment of AI-generated code presents several challenges and controversies. For instance, Johnson et al. (2023) reported that a major financial institution identified critical security vulnerabilities in its trading platform in 2022, which were traced back to segments of AI-generated code. This incident highlights the necessity for rigorous review and testing of AI-generated code, especially in sensitive applications.

The ethical implications of AI-generated code have also sparked considerable debate. Advocates argue that AI-assisted coding democratizes software development and enables developers to concentrate on more complex problem-solving (Smith, 2024). Conversely, critics express concerns regarding the potential for AI to perpetuate biases, introduce undetected errors, and violate intellectual property rights (Brown, 2023).

In response to these issues, recent government initiatives have aimed to enhance transparency in AI-generated content including code. The European Union's AI Act (2021)—currently under review— mandates the disclosure of AI-generated content in certain high-risk applications (European Commission, 2023). Similarly, the National AI Initiative Act of 2020 in the United States emphasizes the importance of explainable AI and transparency in AI systems (U.S. Congress, 2020).

As these regulatory frameworks continue to develop, the capability to detect and analyze AI-generated code becomes essential for compliance. Thus, organizations must implement robust tools and processes to identify, track, and document AI usage throughout their software development lifecycle.

**Need for Transparency in AI-Generated Code**

Transparency in AI-generated code encompasses multiple key dimensions for ensuring the responsible and effective implementation of AI in software development. These dimensions can be categorized as transparency in disclosure, awareness and understanding, and explainability.

**Transparency as Disclosure**

Transparency as disclosure involves the basic acknowledgment and documentation of AI involvement in code generation. This level of transparency is essential for regulatory compliance, intellectual property management, and ethical considerations. As governments and industry bodies develop guidelines for AI use, organizations must clarify and report the extent of AI-generated code in their software products.

**Transparency as Awareness and Understanding**

Beyond mere disclosure, developers and organizations need to possess a deeper awareness and understanding of AI-generated code, i.e., understanding the underlying models and algorithms used in code generation, in addition to the potential limitations and biases inherent in these systems. This understanding is essential for effective code review, maintenance, and troubleshooting.

**Transparency as Explainability**

The most advanced form of transparency is the explainability of AI-generated code. This entails not only recognizing code as AI-generated but also elucidating the reasoning and decision-making processes that produced specific code outputs. The application of explainable AI in code generation can offer insights into the rationale behind certain coding decisions, which will aid developers in evaluating the appropriateness and reliability of the generated code.

**Challenges of Hidden Functionality**

One of the most significant challenges in ensuring transparency in AI-generated code is the potential for hidden functionalities or "Easter eggs." AI systems, especially large language models (LLMs), can generate code with unexpected or unintended behaviors (Zhang et al., 2024). These hidden features may not be immediately apparent but can potentially introduce security vulnerabilities, performance issues, or unethical behaviors into software systems.

The issue of hidden functionality is not exclusive to AI-generated code. Human-written code has also been susceptible to the intentional or unintentional introduction of backdoors, malicious code, or latent bugs. However, the scale and complexity of AI-generated code exacerbate these risks. AI systems may introduce subtle patterns or dependencies that human reviewers cannot easily detect, which potentially induces systemic vulnerabilities across multiple applications (Lee & Park, 2023).

To address these challenges, advanced tools and methodologies need to be developed for code analysis that surpass traditional static and dynamic analysis techniques. These tools should be able to detect potential hidden functionalities, assess code quality, and verify the intended behavior of AI-generated code.

**Code Encryption and Decryption: A Fundamental Paradigm for Transparency**

The challenge of ensuring transparency in AI-generated code can be conceptualized as a problem of code encryption and decryption. Here, "encryption" refers to the inherent complexity and opacity of AI-generated code, whereas "decryption" represents the process of analyzing and understanding this code.

AI models, especially the large-language models used for code generation, can be likened to complex encryption algorithms that transform input prompts and training data into code output. The internal mechanisms of these models are typically opaque, even to their creators, rendering the generated code effectively "encrypted" in terms of its provenance and reasoning.

To achieve true transparency, sophisticated "decryption" tools are required. These tools should analyze AI-generated code and provide insights into its origin, functionality, and potential risks. They should possess the following capabilities:

1. Detection of AI-generated code segments within extensive codebases.
2. Analysis of the structure and patterns of AI-generated code to deduce the underlying model and generation process.
3. Identification of potential vulnerabilities, inefficiencies, or unintended behaviors in the generated code.
4. Provision of explanations for the code functionality and design choices.

However, the development of such decryption tools poses significant technical challenges, which demands advancements in the following areas:

- Machine learning techniques for code analysis.
- Natural language processing to interpret code comments and documentation.
- Program synthesis and verification.
- Application of explainable AI techniques to code generation models.

By conceptualizing the transparency issue in terms of encryption and decryption, we underscore the necessity for a new generation of tools and methodologies specifically designed to analyze and comprehend AI-generated code. These tools are crucial for ensuring the safe and responsible deployment of AI in software development.

**Market Incentives and Disincentives for Transparency**

The implementation of transparency measures for AI-generated code is shaped by competing market forces that create both incentives and disincentives for organizations.

*Incentives for Transparency*

- Regulatory Compliance: As governments and industry bodies formulate regulations governing AI usage, transparency will become a legal mandate in numerous jurisdictions. Organizations that proactively adopt transparency measures will be more adequately equipped to adhere to these emerging regulations.

- Trust and Reputation: Employing transparent practices in AI usage can strengthen the reputation of an organization, which enhances trust among customers, partners, and investors. This is especially valuable in sectors where reliability and ethical considerations are critical.

- Risk Mitigation: Transparency facilitates improved risk management by enabling organizations to detect and rectify potential issues in AI-generated code before they result in security breaches, performance setbacks, or ethical breaches.

- Collaboration and Innovation: Open and transparent practices can encourage collaboration within the developer community, which accelerates innovation and the enhancement of AI coding tools.

- Quality Assurance: Transparency supports more effective code review and testing processes that will lead to the development of higher quality software products.

### *Disincentives for Transparency*

- Competitive Advantage: Organizations may perceive their use of AI in code generation as a competitive edge and may be hesitant to share details that could potentially benefit their competitors.

- Intellectual Property Concerns: There is a concern that transparency could jeopardize intellectual property or expose trade secrets embedded within the AI-generated code.

- Liability Risks: Increased transparency could lead organizations to face greater liability if flaws are found in their AI-generated code.

- Complexity and Cost: The implementation of comprehensive transparency measures can be intricate and expensive, which might deter particularly smaller organizations.

- Performance Concerns: There are apprehensions that transparency mandates could decelerate development processes or degrade the performance of AI coding tools.

Balancing these competing incentives and disincentives is crucial for the widespread adoption of transparency measures in AI-generated code. Policymakers and industry leaders must collaborate to establish frameworks that promote transparency while addressing the legitimate concerns of organizations.

### Challenges of Growing Complexity in Code and Data

The increasing prevalence of AI-generated code is contributing to the overall trend of increasing complexity in software systems. This complexity poses significant challenges for transparency, maintainability, and system reliability.

### *Exponential Growth in Code Volume*

AI-assisted development typically leads to the rapid production of vast amounts of code. Although this can expedite development timelines, it also creates codebases that are increasingly challenging for human developers to completely understand. Li et al. (2023) observed that projects utilizing AI-assisted coding tools experienced a 150% increase in code volume compared to projects without such tools.

### *Increased Interdependencies*

AI-generated code frequently introduces complex interdependencies between various system components. These intricate connections complicate the prediction of the overall effects of modifications or the isolation and resolution of bugs. Chen and Wang (2024) demonstrated that AI-generated code tends to exhibit 30% more intermodule dependencies than human-written code.

### *Data Complexity*

The effectiveness of AI code generation is heavily dependent on the quality and quantity of the training data. With the growing size and complexity of datasets, ensuring their integrity, relevance, and freedom from bias becomes increasingly challenging. A 2023 survey by the AI Ethics Institute revealed that 68% of organizations encounter difficulties in managing and curating the data used to train their AI coding tools.

### *Opacity of AI Models*

Neural networks and machine learning models employed in AI code generation are often characterized as "black boxes" owing to their internal decision-making processes being difficult to interpret. This opacity complicates efforts to explain why specific code is generated or to verify its correctness. Zhang et al. (2024) observed that only 15% of developers are confident in their ability to explain the reasoning behind segments of AI-generated code.

### *Evolving Technology Landscape*

The rapid pace of advancements in AI technology means that tools and techniques for code generation are continually evolving. This dynamic environment challenges the ability of transparency and explainability measures to maintain pace. TechTrends (2023) reported that the average lifespan of AI coding tools before significant updates or replacements is only 18 months.

These challenges highlight the necessity for innovative approaches to manage and understand increasingly complex software systems. As AI assumes a larger role in code generation, formulation of effective strategies for transparency and explainability becomes not only a technical challenge but also essential for ensuring the long-term viability and trustworthiness of software systems.

**Proposed Solutions: Hierarchical Human-AI Hybrid Software Analysis**

To address the growing complexity and opacity of AI-generated code, we propose a hierarchical approach to software analysis that combines human expertise with AI-powered tools. This hybrid model aims to leverage the strengths of both human cognition and machine intelligence to enhance transparency, functionality analysis, and overall code quality.

 Level 1: Automated AI Analysis

The foundational layer of this hierarchical approach comprises automated AI-powered analytical tools designed to:

1. Detect and classify AI-generated code segments within larger codebases
2. Conduct static and dynamic analysis for identifying potential security vulnerabilities, performance issues, and code quality problems
3. Generate initial explanations and documentation for AI-generated code
4. Flag areas of high complexity or potential hidden functionality for further review

 Level 2: AI-Assisted Human Review

Based on automated analysis, the second level involves human developers utilizing AI-powered assistive tools to review and comprehend code. This level includes:

1. Interactive visualization tools that assist developers in navigating complex code structures and dependencies.
2. AI-generated summaries and explanations of code functionality specific to various levels of expertise.
3. Intelligent code comparison tools that emphasize differences between human-written and AI-generated code.
4. Automated test generation and execution to verify code behavior.

 Level 3: Collaborative Human–AI Problem Solving

At the highest level, human experts collaborate with AI systems to address complex issues and execute high-level decisions regarding code architecture and functionality. This involves:

1. AI-powered scenario analysis to predict the long-term impacts of code changes
2. Interactive debugging sessions where AI assistants support human developers in tracing complex issues

3. Collaborative design sessions where AI proposes alternative implementations for human consideration
4. Continuous learning systems that improve AI analysis based on human feedback and decisions

### *Benefits of Hierarchical Approach*

This hierarchical approach offers several advantages:

- **Scalability**: Automated tools manage large volumes of code, whereas human expertise concentrates on critical decision-making and complex problem-solving.
- **Adaptability**: The system can evolve as AI capabilities continue to improve and higher-level tasks are gradually automated.
- **Transparency**: Involving humans at multiple levels ensures a degree of explainability and accountability.
- **Skill Enhancement**: Developers can improve their understanding of complex systems through AI explanations and suggestions.

### *Implementation Challenges*

Implementing this hierarchical approach requires overcoming several challenges:

1. Tool Development: Developing effective AI-powered analysis and assistance tools that integrate seamlessly with existing development workflows.
2. Standards and Protocols: Establishing industry-wide standards for code transparency and AI-human collaboration in software development.
3. Training and Adoption: Educating developers and organizations on how to effectively utilize and trust AI-assisted analysis tools.
4. Ethical Considerations: Ensuring that the AI systems used in code analysis adhere to the ethical guidelines and do not perpetuate biases.

By addressing these challenges and implementing a hierarchical human-AI hybrid approach to software analysis, we can establish a foundation for more transparent, reliable, and understandable AI-generated code. This approach satisfies the increasing demand for explainable AI in critical systems and paves the stage for responsible innovation in software development.

### Long-Term Implications: AI Generation of AI and the Path to AGI

Considering the future of AI-generated codes, it is essential to examine the potential long-term consequences, particularly the possibility of AI systems generating and improving their own codes. The concept of AI-generated AI is no longer only a science fiction scenario. Tools such as Auto-Tune, which employ LLMs to enhance the performance of the LLMs themselves, are in pursuit of recursive self-improvement. At the time of writing this paper, such auto-self-improving AI is minimally effective and still requires significant human intervention for editing, debugging, and creative input, which limits its utility.

However, these tools are continuously improving because of the efforts of numerous skilled developers and substantial investment. It is reasonable to anticipate that autonomous self-improving AI software will become increasingly effective and eventually achieve complete autonomy with profound implications. This self-improving capability could lead to rapid advancements in AI technology, ultimately culminating in the development of AGI.

Several issues related to these trends highlight the importance of transparency in AI-generated code for maintaining human safety in an era of self-improving AI codes:

Control and Alignment: As self-improving AI systems can augment their capabilities by their self, ensuring that they remain aligned with human values and goals becomes a critical challenge.
Transparency and Explainability: As AI systems become more complex and self-modifying, it becomes increasingly challenging to maintain transparency and explainability. This could lead to the emergence of "black box" systems that are opaque even to their creators.
Ethical Considerations: The development of highly advanced AI systems raises profound ethical questions regarding the nature of intelligence, consciousness, and the role of AI in society.

 **Path to AGI: Implications for Code Transparency**

The development of AGI through self-improving AI systems poses significant implications for code transparency and safety.

1. Increased Complexity: AGI systems are likely to be orders of magnitude more complex than current AI systems, which renders traditional approaches to code analysis and transparency inadequate.
2. Dynamic Code Analysis: As AI systems become more dynamic and self-modifying, static code analysis must be supplemented with advanced dynamic analysis techniques capable of real-time monitoring and interpretation of AI behavior.
3. Meta-Level Transparency: Understanding the code itself as well as the processes by which AI systems generate and modify code will be crucial for ensuring safety and alignment.
4. Ethical Frameworks: Robust ethical frameworks and governance structures must be developed to guide the development and deployment of increasingly autonomous AI systems.

 *Convergence of Code and Biology*

As we explore the future of AI-generated codes, the growing convergence between computer science and biology needs further recognition. This intersection has significant implications for our understanding of both fields and the potential applications of AI. Biology can be viewed as a pattern of matter and energy, where the distinction between animate and inanimate matter lies in the information that enables life to function as a viable living system. There is substantial ambiguity in the boundaries between biomatter and information concerning learning, cultural knowledge, and language. Information itself defines our humanity, biology, culture, and technologies. If we consider this as a form of code, we can assert that everything important to humans is the specific code underlying the things we value.

### *Code as the Foundation of Life*

The concept that "everything is code" extends beyond computer systems to biological systems. DNA is the blueprint of life and can be viewed as a biological code that provides information and instructions to living organisms. This perspective is not new; it was explored in depth by John C. Lilly in his 1968 work "Programming and Metaprogramming in the Human Biocomputer" (Lilly, 1968).

More recently, researchers have examined the parallels between computer code and biological systems. For example, the field of synthetic biology treats DNA as a programmable material, which enables scientists to "code" new biological functions (Endy, 2005).

### *AI in Biosciences: Potential and Risks*

The application of AI in the biosciences creates new possibilities for understanding and potentially modifying biological systems. AI algorithms are already used to predict protein structures, design new drugs, and analyze complex biological datasets (Senior et al., 2020).

However, this convergence raises concerns regarding the potential for AI to "hack" biological systems, including the human nervous system. Although current technology is far from achieving this capability, the theoretical possibility signifies the importance of ethical considerations and safety measures in AI development.

### *Implications for Transparency and Safety*

The potential of AI to interact with and modify biological systems introduces additional complexity to issues of code transparency and safety. As AI systems become increasingly involved in biological research and applications, ensuring transparency and safety extends beyond software integrity to encompass biological and ecological safety.

This convergence underscores the necessity for interdisciplinary approaches to AI safety and ethics, involving not only computer scientists and AI researchers but also biologists, ethicists, and policymakers.

## Dual Nature of AI Advancement

In considering the future of AI-generated codes and their potential to influence technological and biological systems, the dual nature of these advancements should be acknowledged. Although significant risks and challenges exist, substantial potential benefits can also be realized.

### *Potential Benefits*

1. Enhanced Human Capabilities: AI systems could augment human intelligence, leading to breakthroughs in science, medicine, and technology.
2. Solving Complex Problems: Advanced AI could help address global challenges such as climate change, disease, and resource scarcity.
3. Personalized Education and Healthcare: AI could enable highly personalized approaches to education and healthcare, which improves outcomes for individuals.
4. Economic Growth: AI-driven automation and innovation could result in significant economic growth and improved quality of life.

### *Mitigating Risks*

Several key steps should be followed to realize these benefits while mitigating risks.

Robust Transparency Measures: Development of advanced tools and methodologies to ensure transparency in AI systems—from code generation to high-level decision-making processes.
Ethical Frameworks: Establishing comprehensive ethical guidelines for AI development and deployment, with particular attention to issues of autonomy, privacy, and human rights.
International Cooperation: Promoting global collaboration on AI safety and ethics to ensure that the advancements benefit humanity in its entirety.
Public Engagement: Engaging communities in discussions related to AI development to ensure that the technology aligns with societal values and needs.
Interdisciplinary Research: Promoting collaboration between computer scientists, biologists, ethicists, and other relevant fields to address the complex challenges posed by advanced AI.

**Conclusions**

The rapid advancement of AI-generated code presents unprecedented opportunities and significant challenges in the field of software engineering. This study explores the importance of ensuring the transparency, safety, and ethical use of AI in code generation to harness its benefits while mitigating potential risks.

The necessity for robust transparency measures in AI-generated code is evident. Transparency is indispensable for maintaining the integrity and reliability of software systems by detecting hidden functionalities and understanding complex interdependencies. The proposed hierarchical human–AI hybrid approach to software analysis offers a promising framework for addressing these challenges, leveraging the strengths of both human expertise and AI capabilities.

In future, the potential for AI to generate and improve its own code raises profound questions regarding the nature of intelligence and the long-term trajectory of AI development. The convergence of computer science and biology highlights the far-reaching implications of these advancements.

A proactive and collaborative approach is necessary to navigate this complex landscape, which involves the following:

1. Investment in research and development of advanced transparency and safety tools for AI-generated code.
2. Establishment of comprehensive ethical frameworks and governance structures for AI development.
3. Promotion of interdisciplinary collaboration to address the multifaceted challenges posed by advanced AI.
4. Engagement in public discourse to ensure that AI development aligns with societal values and needs.

By inculcating these steps, we can work toward a future in which AI-generated code enhances human capabilities, drives innovation, and contributes to solving global challenges while maintaining the safety, transparency, and ethical standards necessary for responsible technological advancement.

The future of code is inextricably linked to that of AI. By prioritizing transparency, safety, and ethical considerations in AI-generated code, we can help shape a future in which technology serves humanity's best interests and opens new frontiers of possibility.